# Ontological determinism, non-locality, quantum equilibrium and post-quantum mechanics

**Maurice Passman**
Adaptive Risk Technology, Ltd.
London, UK
adaptiverisktechnology@gmail.com

**Philip V. Fellman**
American Military University
Charles Town, WV
Shirogitsune@gmail.com

**Jonathan Vos Post**
Computer Futures
Altadena, CA
Jvospost3@gmail.com

**Avishai Passman**
Adaptive Risk Technology, Ltd.
London, UK
adaptiverisktechnology@gmail.com

**Jack Sarfatti**
Internet Science Education Project
San Francisco, CA and London, UK
jacksarfatti@icloud.com

## Abstract

In this paper, we extend our previous discussion on ontological determinism, non-locality and quantum mechanics to that of Sarfatti's post-quantum mechanics (PQM) perspective. We examine the nature of quantum equilibrium/non-equilibrium and uncertainty following Sarfatti's description of this theoretical development, which serves "to extend the statistical linear unitary quantum mechanics for closed systems to a locally-retrocausal, non-statistical, non-linear, non-unitary theory for open systems." [20,21] We discuss how the Bohmian quantum potential has a dependence upon the position of its Bell 'beable' and how Complexity mathematics "describes the self-organizing feedback between the quantum potential and its beable allowing nonlocal communication." [20,21]

# 1 Introduction

Our previous paper [1] reviewed the "measurement" problem in quantum mechanics. It followed our earlier paper, "The Fundamental Importance of Discourse in Theoretical Physics" [2], which paid especial attention to the arguments of John S. Bell, particularly those dealing with Bohr's "debate" with Einstein on the language and meaning of quantum mechanics. This previous paper entered into a more technical treatment of non-locality as well as demonstrating that even where quantum mechanics is deterministic, this is not an ontological necessity; this treatment came from a point of view that can be called a Bohmian perspective. This perspective, named after its inventor, David Bohm, takes the view that the positions of particles constitute the primitive variables and thus the primary ontology. These particles are therefore 'beables' in Bell's sense.

The formulation of this perspective does not involve the notion of quantum observables, as given by self-adjoint operators [3], however, the predictions of this perspective concerning the results of a quantum experiment (provided it is assumed that prior to the experiment the positions of the particles in the system are distributed according to Born's Law) are the same as the predictions of a quantum formalism. At the heart of this ontology is Smoluchowski's question [4] 'How can randomness arise, if all events are reducible to deterministic laws of nature?'. Within the Bohmian perspective, one can resort to Boltzmann-Gibbs ensemble definitions of typicality and Quantum Equilibrium [5].

In this paper, we wish to follow Sarfatti's framework and in his words, "extend our discussion to how the Bohmian quantum potential has a dependence upon the position of its Bell 'beable' and how Complexity mathematics describes the self-organizing feedback between the quantum potential and its beable allowing nonlocal communication. The first stage of this discussion is to examine quantum equilibrium and thus by extension quantum non-equilibrium." [20,21]

The main reason that the Bohmian perspective was initially so compelling to us was there were ontological similarities to other areas of our work, namely, the search for an explanation of why fractal structures seemed to be everywhere within nature. Fundamental to this search was Bak's Self Organised Criticality (SOC), far-from-equilibrium systems, and the notion that smaller scale mechanisms were the drivers for larger scale observable behaviors. SOC can be defined as scale invariance without external tuning of a control parameter, but with all the features of the critical point of an ordinary phase transition and, in particular, long range spatiotemporal correlations. Fractal structures are associated with power laws and scaling and therefore can possibly be used as a predictive instrument for looking at 'real life' systems such as financial markets [6,7,8,9]. However, there appears to be a major problem with this notion: it doesn't work. What we see in real systems are crashes that are more frequent and more extreme than we would expect (Sornette's 'outliers' See: [9]). Together with this unexpected deviation from predicted values, there is some confusion within the area of study of financial 'crash' systems. Many of the publications under the name of Complexity have little to do with the mathematical techniques of Complexity Science [10,11] and more to do with grafting on the Complexity 'label' to micro- and macro-economic studies [12]. Our conclusion drawn from our previous work undertaken in Quantum Mechanics and Complex systems (see for example [13]), is that there must be a secondary driving mechanism underlying the SOC behavior we see in these real systems: as Sarfatti has recognized, that of "Frohlich Pumping." [20,21]

In this current paper, therefore, we wish to lay the framework for a formalism that is inclusive of Quantum Equilibrium and Quantum Disequilibrium but also to allow this framework to be a stepping stone which will enable us to garner additional tools to study larger scale 'real life' systems: particularly those of financial institutions. The reason for this particular approach is that global financial crises have precipitated an increasing awareness of the necessity of obtaining a systematic perspective of financial stability. Therefore, a fundamental question for policymakers is how to shape policy in the light of emerging financial risks. For example, how do financial institutions, such as banks shape systematic risk and how does their concentration, connectivity, cross-capitalization and diversification contribute to stability? Our starting point will be to examine and develop such a framework from the basis of a "pumped, non-equilibrium, Frohlich coherent system" [20,21] that, in turn, leads to characteristic fingerprint behaviors.

## 2. Quantum Equilibrium

### 1. Issues with the Quantum Equilibrium Hypothesis

The Quantum Equilibrium Hypothesis (QEH) can be stated as follows. For an ensemble of identical systems, each having a wave function, the typical empirical distribution of the configurations of the particles is given approximately by ρ= 2. This is Born's statistical law. This property had already been emphasized by de Broglie in 1927 and was later formalized and called equivariance by Durr et al. [3], who used it to establish the typicality of empirical statistics given by the quantum equilibrium distribution. The notion of equivariance is a natural generalization of the stationarity of a distribution in statistical mechanics and dynamical systems theory [3]. Quantum equilibrium invokes the typical behavior of particle positions given the wave function. This implies that non-equilibrium resides within the wave function. The temptation is therefore to seek, much like in classical physics, a measure of typicality. This effort is resolved by use of a Boltzmann-Gibbs ensemble in first seeking a measure with which a statistical hypothesis about the typical empirical distribution, over an ensemble of identical systems, can be formed. The QEH was presented in Bohm's original papers of 1952 as derivable from statistical-mechanical arguments. This approach was further supported by the work of Vigier and Bohm's paper of 1954, in which they introduced stochastic fluid fluctuations that drive a process of asymptotic relaxation from quantum non-equilibrium to quantum equilibrium [14].

An examination of the QEH leads to a number of insights:

1. **The Bohmian Mechanics particle trajectory may be undefined.**

The equation for an *N*-particle system where $Q_k(t)$ is the k'th particle trajectory within a Bohmian framework is written as [4]:

$$\frac{dQ_k}{dt} = \frac{\hbar}{m_k} I_k(Q,t)$$

Note that this has the wave function in the denominator. The equation breaks down when the wave function is zero. Again, one may have recourse to typicality and that the wave function must be differentiable for the classical solutions of the Schrodinger equation. Therefore, the existence and uniqueness of Bohmian trajectories for all times and 'almost all' initial conditions is where this 'almost all' refers to the quantum equilibrium distribution.

2. **If Quantum Disequilibrium were possible so would superluminal signalling: Entanglement**
Consider the EPR experiment example from our previous papers [1,4] we found that the 'statistical' outcomes are the same no matter if the measurement takes place on the left or right hand magnet. Take a general entangled two-particle state:

$$\psi = a|\uparrow\rangle_1|\downarrow\rangle_2 + b|\downarrow\rangle_1|\uparrow\rangle_2 + c|\downarrow\rangle_1|\downarrow\rangle_2 + d|\uparrow\rangle_1|\uparrow\rangle_2$$

with $|a|^2 + |b|^2 + |c|^2 + |d|^2 = 1$.

The probability of getting the spin value $\left|\uparrow\right\rangle_2$ at the right magnet is $|b|^2+|d|^2$. Now undertake a measurement at the left magnet in an arbitrary direction $\gamma$, where $\gamma$ is the angle between the chosen direction and the z-direction. We can now write the entangled equation in the $\gamma$-basis:

$$\left|\uparrow\right\rangle_1 = \mathbf{i}_1 \cos\gamma + \mathbf{j}_i \sin\gamma \quad \text{and} \quad \left|\downarrow\right\rangle_1 = -\mathbf{i}_1 \sin\gamma + \mathbf{j}_i \cos\gamma$$

We can rewrite the state $\psi$ as:

$$\psi = \mathbf{i}_1\left[\cos\gamma(a\left|\downarrow\right\rangle_2 + d\left|\uparrow\right\rangle_2) - \sin\gamma(b\left|\uparrow\right\rangle_2 + c\left|\downarrow\right\rangle_2)\right] + \mathbf{j}_1\left[\sin\gamma(a\left|\downarrow\right\rangle_2 + d\left|\uparrow\right\rangle_2) + \cos\gamma(b\left|\uparrow\right\rangle_2 + c\left|\downarrow\right\rangle_2)\right]$$

$$\approx \psi_{\mathbf{i}_1} + \psi_{\mathbf{j}_1}$$

Therefore $\left\|\psi_{\mathbf{i}_1}\right\|^2$ and $\left\|\psi_{\mathbf{j}_1}\right\|^2$ are the probabilities for the outcomes spin up or spin down when measuring first at the left magnet at an arbitrary angle $\gamma$. The collapsed wave function will either be $\dfrac{\psi_{\mathbf{i}_1}}{\left\|\psi_{\mathbf{i}_1}\right\|}$ or $\dfrac{\psi_{\mathbf{j}_1}}{\left\|\psi_{\mathbf{j}_1}\right\|}$ depending upon the outcome at the left-hand magnet. The effect of this measurement on the probability for the outcome at the right magnet is as follows.

Calculating the equilibrium probability for the spin up wave function $\left|\uparrow\right\rangle_2$, for the collapsed wave function $\dfrac{\psi_{\mathbf{i}_1}}{\left\|\psi_{\mathbf{i}_1}\right\|}$, we obtain the probability:

$$\frac{\left\|\mathbf{i}_1\left|\uparrow\right\rangle_2 (d\cos\gamma - b\sin\gamma)\right\|^2}{\left\|\psi_{\mathbf{i}_1}\right\|^2}$$

And for $\dfrac{\psi_{\mathbf{j}_1}}{\left\|\psi_{\mathbf{j}_1}\right\|}$, we obtain:

$$\frac{\left\|\mathbf{j}_1\left|\uparrow\right\rangle_2 (b\cos\gamma - d\sin\gamma)\right\|^2}{\left\|\psi_{\mathbf{j}_1}\right\|^2}$$

The outcome for the spin up case for the right-hand magnet is:

$$\left\|\psi_{\mathbf{i}_1}\right\|^2 \frac{\left\|\mathbf{i}_1\left|\uparrow\right\rangle_2 (d\cos\gamma - b\sin\gamma)\right\|^2}{\left\|\psi_{\mathbf{i}_1}\right\|^2} + \left\|\psi_{\mathbf{j}_1}\right\|^2 \frac{\left\|\mathbf{j}_1\left|\uparrow\right\rangle_2 (b\cos\gamma - d\sin\gamma)\right\|^2}{\left\|\psi_{\mathbf{j}_1}\right\|^2} = |b|^2 + |d|^2$$

Therefore the 'statistical' outcomes are the same no matter if the measurement takes place on the left or right hand magnet. The experiment demonstrated that Bohmian QM is non-local. The key properties that were used in the proof were the commutation of spin observables in that we can infer the probability for the outcome of the particle through SGM-R by summing the joint probability over the possible values of the outcome from SGM-L. The commutation of spin observables is an expression that local operations can be performed i.e. that the left and right magnets are decoupled but there is no effect on the statistics of the outcomes on the right; they are the same whether or not a measurement on the left takes place first.

### 3. Our universe is atypical

A typical universe is a quantum equilibrium universe [4]. But our universe is atypical. The second law of thermodynamics seems to justify the atypical initial conditions which are synonymous with non-equilibrium conditions. Thus, in practice, we observe non-equilibrium, far-from-equilibrium, out-of-equilibrium as well as equilibrium behaviors [15].

### 2. Retrocausality

Bohm and Hiley in their 'The Undivided Universe' [16] first brought up the notion of retrocausality: "If there is a nonlocal connection of the kind implied by our guidance conditions, then it follows that, for example, point a and point b instantaneously affect each other. But if the theory is covariant, there should be similar instantaneous connections in every Lorentz frame.... It would then be possible for A acting at a to affect its own past."[17].

As Sarfatti has noted in his examination of retrocausality, "Sutherland introduced, through a number of papers [18,19], a locally retrocausal relativistic action-reaction post-Bohmian Lagrangian. This eliminates configuration space for entanglement and is non-linear, non-unitary and non-statistical." [20,21] Within this formalism the linear, unitary statistical Born rule orthodox quantum theory is a limiting case of the more general theory. Sarfatti labelled this formalism as "Post-Quantum Mechanical (PQM)." It's defining characteristics are:

- It is non-statistical, there is no Born rule: God does not play dice because of local retrocausality - quantum mechanics is a limiting statistical, retarded causality only sub-theory when the action-reaction is negligible and the Born rule is assumed as a new axiom connecting advanced and retarded pilot waves. This allows the integration away of future causes of present effects (see below) leaving only past causes of those same present effects when we limit our attention only to von Neumann projection operator strong measurements;
- It eliminates the need for configuration space for spatio-temporally separated yet entangled particles;
- It is non-unitary because of action-reaction where classical matter is a source for the advanced and retarded Bohm-Aharonov weak measurement pilot waves (the $w$ suffix in the formalism below);
- It is nonlinear. [20,21]

Again following Sarfatti, we find from Sutherland [18], "the free-particle spin 1⁄2 Dirac Equation for the retarded history pilot wave <x|i> is given by the following. Note that the right-hand side of the equation is the back-reaction *of the particles on their wave*:

$$(i\gamma^\alpha \partial_\alpha - m)\langle x|i\rangle = \sigma_0 \left( u_\alpha \pm \frac{j_{\alpha w}}{\rho_{0w}} \right) \langle x|i\rangle$$

$$\sigma_0 = \frac{1}{u^0} \,{}^3\!\left(\vec{x} - \vec{x}_p()\right)$$

$$\rho_{0w} = \sqrt{(j_w j_w)}$$

$$j(x)_w = \frac{\langle f|x\rangle \langle x|i\rangle}{\langle f|i\rangle}$$

Sutherland's post-quantum weak measurement action-reaction "factor" when set to zero is de Broglie's guidance constraint:

$$\left( u \pm \frac{j_w}{\rho_{0w}} \right)$$

Where u is the classical h-independent particle/beable 4-velocity and jαw is the Aharonov weak value destiny/history pilot wave current density. Again, from Sutherland [18,19], the particle/beable equation of motion for the back reaction of the wave function upon the particles is:

$$\frac{d(_{0w}u)}{d\tau} \rho \pm \frac{\alpha \partial_{0w}}{\partial x} + u \left( \rho \frac{\partial j_w}{\partial x} - \beta_\alpha \frac{\partial \dot{j}_w}{\partial x} \right)$$

We therefore have two non-linear equations that form an adaptive, self-organising feed-back loop type process between the 'beable' and the wave function. The action-reaction term must be assumed to be zero to justify integrating over the back from the future <f| Aharonov-Dirac destiny "bras".[20,21]

There would be a non-random locally-decodable keyless signal from the actual future <f| when the action-reaction term was sufficiently large not equal to zero:

$$J_{PQM}(x) \equiv_\mu \frac{\langle f|x\rangle \hat{j} \langle x|i\rangle}{\langle f|i\rangle}$$

$$J_{QM}(x) = \int_f |\langle f|i\rangle|^2 J_{PQM}(x)_\mu = \langle i|x\rangle \hat{j} \langle x|i\rangle$$

$$\int_f |f\rangle\langle f| = 1$$

$$\langle J_{QM}\rangle \equiv \int_x \langle i|x\rangle \hat{j} \langle x|i\rangle \equiv \langle i|\hat{j}|i\rangle$$

To be clear: we see that 'God does not play dice' in the PQM formalism but does 'play dice' in the QM formalism i.e:

$$J_{PQM}(x) =_\mu \frac{\langle f|x\rangle \hat{j} \langle x|i\rangle}{\langle f|i\rangle}$$

$$J_{QM}(x) = \int_f |\langle f|i\rangle|^2 J_{PQM}(x)_\mu = \langle i|x\rangle \hat{j} \langle x|i\rangle$$

$$\int_f |f\rangle\langle f| = 1$$

$$\langle J_{QM}\rangle = \int_x \langle i|x\rangle \hat{j} \langle x|i\rangle = \langle i|\hat{j}|i\rangle$$

Note that the second line uses the QEH, i.e. the conditional Born Rule Axiom <f|i>2. Thus the last line above "corresponds to Stapp's orthodox statistical quantum theory in which the initial state evolves from past to future in accord with the thermodynamic Arrow of Time and the accelerated expansion of our classical relativity light-signal limited observable universe bounded by observer-dependent past-particle cosmological horizons encoding |i> and the future de Sitter event horizons encoding <f|" [21]. Thus, in the Bohm-Hiley formalism, the wave function modified the particle (quantum potential), however, there is no feedback of the particle to the wave function. In Sutherland's formalism, the action-reaction un-hides the beables. Their path therefore no longer needs to coincide with the pilot wave 'stress lines' [21].

## 3. Frohlich Pumping

As mentioned earlier in this paper, one of the drivers of our research was the search for a predictive mechanism for 'real life', open systems such as financial markets. We wished to improve the situation where most of the behavioral frameworks that have attempted to describe such systems were in their essence endogenous. In such systems, there was no real understanding of the nature of exogenous effects and how these exogenous effects integrated themselves into the endogenous behavioral characteristics of the agents that made up these behavioral/simulation models. The real systems displayed fingerprint behaviors of dissipative, fractal and self-organizing criticality, but also exhibited evidence of extremal behaviors, i.e. outliers to the usual 'crash' behaviors. Examination of the data from a number of studies (See for example [22] and references within) suggested to us that what in every-day occurrence was modeled as a system in equilibrium [23] was actually being 'stoked' or 'pumped' to then exhibit not only crashes but crashes that were more extreme than and deviated from those predicted by power law behavior.

Sarfatti has linked the back-reaction of PQM to Frohlich coherence specifically as he was searching for a mechanism that would explain the driving mechanism behind PQM [21]. Sarfatti has also produced a toy (i.e. not fully developed) mathematical formalism for a generalised Frohlich, externally pumped, many-particle system [21]. What was interesting from our perspective was that here was the suggestion for driver for the SOC type behaviors we were observing but also a mathematical formalism that centred on the quantum-mechanical *statistically irreversible* thermodynamics of these open systems, and the informational characteristics of the phenomena, that mirrored the dynamics and characteristics of the real systems that we were examining. A further paper is currently in progress, using the scalar equilibrium economic model [23] as a basis to investigate this relationship more fully.

## 3. Summary

The main contribution of this paper has been to extend our previous discussion of ontological determinism, non-locality and quantum mechanics to that of a post-quantum mechanics (PQM) perspective. We examine the nature of quantum equilibrium/non-equilibrium and uncertainty to extend the statistical linear unitary quantum mechanics for closed systems to a locally-retrocausal, non-statistical, non-linear, non-unitary theory for open systems. Our motivation for this exercise was based upon the wish to examine and scale the fundamental equilibrium and non-equilibrium mechanisms and to observe if there were any consistencies within this methodology that could be applied to practical real-life problems. We have noted a number of issues concerned with the QEH that we wish to further develop and explore, particularly with reference to transluminal signalling. The notion of typicality that is fundamental to QEH has led us to further explore financial systems models that do not fully represent real life behavior. The results of this exploration, using the equilibrium based Skander bank capitalization model as a basis, will be presented within a further paper.